\documentclass{myarxiv}

\usepackage{amsmath}
\usepackage{amssymb}
\usepackage{mathrsfs}
\usepackage{makecell}
\usepackage{url}
\usepackage[english]{babel}
\usepackage{wrapfig}
\usepackage{array}
\usepackage{graphicx}
\usepackage{multirow}
\usepackage{bigstrut}
\usepackage{booktabs}
\usepackage{txfonts}
\usepackage{color}

\newcommand{\ii}{\textnormal{i}}
\newcommand{\ee}{\textnormal{e}}
\newcommand{\X}[1]{$\textnormal{#1}_\textnormal{D}$}
\newcommand{\D}[1]{$\widetilde{\textnormal{#1}}$}
\newcommand{\DX}[1]{$\widetilde{\textnormal{#1}}_\textnormal{D}$}
\newcommand{\ytab}[2]{
  \begin{tabular}{@{}c@{}c@{}}$#1$ & $\,\,(\sigma_2/\sigma_1\ll 1)$ \\ $#2$ & $\,\,(\sigma_2/\sigma_1\gg 1)$\end{tabular}
}
\newcommand{\red}{\color[RGB]{192,0,0}}

\def\ack{\section*{Acknowledgements}
  \addtocontents{toc}{\protect\vspace{6pt}}
  \addcontentsline{toc}{section}{Acknowledgements}
}
\renewcommand{\theenumi}{\roman{enumi}}

\begin{document}

\begin{frontmatter}

\begin{minipage}{140mm}
\title{
Fourier-based schemes with modified Green operator for computing the electrical response of heterogeneous media with accurate local fields
}

\author[mines]{Fran\c{c}ois Willot*}
\ead{francois.willot@ensmp.fr}
\author[mines]{Bassam Abdallah}
\author[dif]{Yves-Patrick Pellegrini}
\address[dif]{CEA, DAM, DIF, F-91272 Arpajon, France.}
\address[mines]{MINES ParisTech, PSL - Research university, CMM - Centre for mathematical \\ morphology,
35, rue St Honor\'e, F-77300 FONTAINEBLEAU, France.}
\end{minipage}


\vspace{1cm}
\begin{minipage}{140mm}
{\red \textit{Nota Bene}: The present document constitutes a `postprint' version of the published paper, in which a few errors in the proofs (on the present pages 2, 5, 15 and 17) have been corrected and an incomplete reference on p.\ 22 has been completed. These minor corrections are marked out in red. Results are unchanged.}
\end{minipage}
%\medskip

\begin{abstract}
A modified Green operator is proposed as an improvement of Fourier-based numerical schemes commonly used for computing the electrical or thermal response of heterogeneous media. Contrary to other methods, the number of iterations necessary to achieve convergence tends to a finite value when the contrast of properties between the phases becomes infinite. Furthermore, it is shown that the method produces much more accurate local fields inside highly conducting and quasi-insulating phases, as well as in the vicinity of phase boundaries. These good properties stem from the discretization of Green's function, which is consistent with the pixel grid while retaining the local nature of the operator that acts on the polarization field. Finally, a fast implementation of the `direct scheme' of Moulinec \emph{et al.} (1994) that allows for parsimonious memory use is proposed.
\end{abstract}

\end{frontmatter}

% ----------------------------------------
\section{Introduction}\label{sec:intro}
In recent years, Fourier-based methods, originally introduced by Moulinec \emph{et al.}~\cite{Moulinec94xx}, have become ubiquitous for computing numerically the properties of composite materials, with applications in domains ranging from linear elasticity~\cite{Willot08}, {\red visco\-plasticity [instead of `thermoplasticity']}~\cite{a2001n}, and crack propagation~\cite{li2011non} to thermal and electrical~\cite{Willot13a,Willot10} and also optical properties~\cite{Willot13b}. The success of the method resides in its ability to cope with arbitrarily complex and often very large microstructures, supplied as segmented images of real materials, for example, {\red multiscale [instead of `multistage']} nanocomposites~\cite{Jean2011large}, austenitic steel~\cite{brenner11}, granular media~\cite{Willot13a} or {\red polycrystals [instead of `polycrystal']}~\cite{Lebensohn09,Rollett10,lebensohn2005study}. This technique allows maps of the local fields to be computed in realistic microstructures. Such fields are representative of the material behavior if the resolution is small enough, and if the system size is large enough, compared with the typical length scale of the heterogeneities. Contrary to finite-element methods (FEM) where matrix pre-conditioning often necessitates additional memory occupation, fast-Fourier-transform (FFT) methods are limited only by the amount of RAM or fast-access computer memory required to store the fields.

The use of an image and of its underlying equispaced grid, however, comes with drawbacks not seen in FEM. First, FFT methods will ultimately be less efficient when dealing with highly porous media such as foams, where voids need to be discretized. Second, interfaces are crudely rendered when using voxel grids, although smoothness can be somewhat recovered by introducing intermediate properties between phases~\cite{Dunant13,Brisard10}. This matter is the most important one for ideal microstructure models where interfaces are completely known; less so when dealing with experimental images where such information is usually absent. Third, the representation of the fields in terms of harmonic functions introduces oscillations around interfaces, which is akin to Gibbs's phenomenon. High-frequency artifacts are conspicuous in many field maps where oscillations are visible. Fourth, the Fourier representation presupposes periodicity; that is, the microstructure is seen as the elementary cell of an infinite, periodic medium. However, finite-size effects associated to periodic boundary conditions are generally smaller than that of uniform boundary conditions used in FEM~\cite{RVE1}.

In the present work, use is made of an alternative discretization of the Green function, leading to a revisit of some previously developped FFT algorithms. Specifically, their performances in terms of accuracy and speed are investigated. Our paper is organized as follows: the numerical problem and FFT algorithms are presented in Secs.\ \ref{sec:prob} and\ \ref{sec:FFTmethods}, respectively. An alternative discretization is introduced in Section \ref{sec:green}. The accuracy of the local fields is investigated in Section \ref{sec:local} and the convergence properties of FFT schemes, using the modified and unmodified Green functions, are studied in Section \ref{sec:results}. Finally, a specific implementation of the FFT method using the modified Green function is proposed in Section \ref{sec:speed}.

% ----------------------------------------
\section{Problem setup and Lippmann-Schwinger's equation}
\label{sec:prob}\label{sec:ls}
This work investigates the numerical computation of the electric field $E_i(\mathbf{x})$ and current $J_i(\mathbf{x})$ ($i=1$, ..., $d$), in a $d$-dimensional cubic domain $\Omega=[-L/2,L/2]^d$ of width $L$ for $d=2$ or $3$. The fields verify~(chapter 2 in~\cite{MiltonBook})
\begin{equation}\label{eq:edp}
  \partial_i J_i(\mathbf{x})=0, \qquad E_i(\mathbf{x})=-\partial_i \Phi(\mathbf{x}),
  \qquad J_i(\mathbf{x})=\sigma_{ij}(\mathbf{x}) E_j(\mathbf{x}),
\end{equation}
where $\Phi(\mathbf{x})$ is the electric potential and $\boldsymbol\sigma(\mathbf{x})$ is the local conductivity tensor of the material phase at point $\mathbf{x}$. Thereafter, for simplicity, all media are locally linear and isotropic so that $\sigma_{ij}=\sigma\delta_{ij}$, with $\boldmath{\sigma}(\mathbf{x})$ a scalar field.  Only binary composite media are considered in this study, in which inclusions have variable conductivity $\sigma_2$, and where conventionally $\sigma_1$=1 in the matrix. Edges of $\Omega$ are aligned with the Cartesian axis of unit vectors $(\mathbf{e}_i)_{1\leq i\leq d}$. Periodic boundary conditions are employed, in the form
\begin{equation}
\mathbf{J}(\mathbf{x})\cdot \mathbf{n}\,-\#,
  \quad \Phi(\mathbf{x}+L\mathbf{e}_i)\equiv \Phi(\mathbf{x})-\overline{E}_iL,
\quad\mathbf{x},\, \mathbf{x}+L\mathbf{e}_i\in \partial\Omega,
\end{equation}
where $-\#$ denotes anti-periodicity, $\mathbf{n}$ is the outer normal along the boundary $\partial\Omega$ of $\Omega$ and $\overline{\mathbf{E}}$ is the applied electric field. They ensure that the current and the electric field verify Equation~(\ref{eq:edp}) along the boundary $\partial\Omega$ of the periodic medium. Note that $\overline{\mathbf{E}}$ represents a macroscopic electric field so that $\langle E_i(\mathbf{x})\rangle = \overline{E}_i$, where $\langle \cdot\rangle$ is the volume average over $\Omega$.

All FFT methods proceed from Lippmann-Schwinger's equation~(\cite{MiltonBook} p.\ 251)
\begin{equation} \label{eq:ls1}
E_i=\overline{E}_i-G^0_{ij}\ast P_j,
\quad P_j=J_j-\sigma^0E_j,
\end{equation}
where $\sigma^0$ is an arbitrary reference conductivity, $\mathbf{P}$ and $\mathbb{G}^0$ are the associated polarization field and Green operator, respectively, and $\ast$ is the convolution product. An equivalent `dual' formulation stems from writing the problem in terms of the electric current as
\begin{equation} \label{eq:ls2}
J_i=\overline{J}_i-H^0_{ij}\ast T_j,
\quad T_j=E_j-\rho^0J_j,
\end{equation}
where $\rho^0=1/\sigma^0$ is the reference resistivity, and $\overline{\mathbf{J}}$ is the prescribed macroscopic current. The Green operator associated to the governing equation for the current reads
\begin{equation}\label{eq:greendual}
H_{ij}^0(\mathbf{x})=\sigma^0\left\lbrace\left[\delta(\mathbf{x})-1\right]\delta_{ij}-\sigma^0G^0_{ij}(\mathbf{x})\right\rbrace,
\end{equation}
where $\delta(\mathbf{x})$ is Dirac's distribution and $\delta_{ij}$ is the Kronecker symbol. Thus, for all $\mathbf{T}$,
\begin{equation}
H_{ij}^0\ast T_j=\sigma^0\left(T_i-\langle T_i\rangle_\Omega-\sigma^0G^0_{ij}\ast T_j\right).
\end{equation}
In particular, $\langle H_{ij}^0\ast T_j\rangle=\langle G^0_{ij}\ast T_j\rangle=0$ and Equation~(\ref{eq:ls2}) enforces $\overline{\mathbf{J}}=\langle \mathbf{J}\rangle$. The FFT algorithms considered in this paper rest on evaluating the convolution product in Equation (\ref{eq:ls1}) or (\ref{eq:ls2}) in the Fourier domain, using FFT libraries.

% ----------------------------------------
\section{FFT methods}
\label{sec:FFTmethods}
Although most of FFT methods have been introduced in the context of elasticity, their adaptation to conductivity problems is straightforward. Hereafter, all FFT algorithms are formulated in this setting.
\label{subsec:FFTalg}
Equation (\ref{eq:ls1}) is the basis of the simplest method, the `direct' scheme~\cite{Moulinec94xx}. Iterations consist in applying the following recursion:
\begin{equation}\label{eq:ls1imp}
\mathbf{E}^{k+1}=\overline{\mathbf{E}}-\mathbb{G}^0\ast \left[(\sigma-\sigma^0)\mathbf{E}^k\right],
\end{equation}
where $\mathbf{E}^k$ is the electric field at iteration $k$.

Over time, refined FFT algorithms with faster convergence properties have been devised, notably the `ac\-cel\-era\-ted'~\cite{Eyre99} and `augmented-Lagrangian'~\cite{Michel01} schemes. Both algorithms can be encapsulated in the formula~\cite{Monchiet12,moulinec13}
\begin{equation}\label{eq:alphabeta}
\mathbf{E}^{k+1}=\mathbf{E}^k+\frac{\sigma^0\left[\overline{\mathbf{E}}-\langle \mathbf{E}^k\rangle-\beta \mathbb{G}^0\ast (\sigma \mathbf{E}^k)\right]-\mathbb{H}^0\ast \mathbf{E}^k}{\alpha(\sigma+\beta\sigma^0)}
\end{equation}
where $\alpha=\beta=1$ for the augmented-Lagrangian scheme and $\alpha=-1/2$, $\beta=-1$ for the `accelerated' one. Our formula differs from Equation~(13) in~\cite{moulinec13} because of a different definition of $\sigma^0$. Another scheme, the so-called `polarization' scheme where $\langle \mathbf{P}\rangle$ is prescribed instead of $\langle \mathbf{E}\rangle$, can be described by an equation similar to (\ref{eq:alphabeta})~\cite{Monchiet12}.

The alternative `variational' algorithm~\cite{Brisard10} relies on two distinct ideas. First, Equation~(\ref{eq:ls1}) is written as:
\begin{equation}\label{eq:var}
\left[(\sigma-\sigma^0)^{-1}\delta(\mathbf{x})\delta_{ij}+G^0_{ij}\right]\ast P_j= \overline{E}_i.
\end{equation}
Upon discretization, this equation is transformed into a linear system $\mathcal{M}\cdot \mathbf{P}=\overline{\mathbf{E}}$, which is solved by conjugate-gradient descent. The operator $\mathcal{M}$ is never computed. Instead, FFTs are used to provide $\mathcal{M}\cdot \mathbf{P}$ for any $\mathbf{P}$, which is sufficient for applying the descent method. Second, the discretization employed amounts to using constant-per-voxel trial polarization fields. This leads to a rule for computing $(\sigma-\sigma^0)^{-1}\mathbf{P}$ on voxels that lie on interfaces, and to a representation of the Green operator as a slowly converging series for which approximations are available~\cite{Brisard12}.

Other FFT methods have been proposed, including an alternative `conjugate-gradient' scheme~\cite{Zeman10,Vondrejc11} different from the variational one, and yet another one in which the convolution product is carried out in the direct space~\cite{yvonnet2012fast}. For conciseness, these and the `polarization' scheme will not be considered further.

The dual formulation (\ref{eq:ls2}) allows one to derive dual algorithms for all FFT methods. For instance, substituting $\mathbf{E}$, $\mathbb{G}^0$, and $\sigma^0$ by $\mathbf{J}$, $\mathbb{H}^0$, and $\rho^0$ in Equation~(\ref{eq:alphabeta}), the dual augmented-Lagrangian scheme reads:
\begin{equation}
\mathbf{J}^{k+1}=\mathbf{J}^k+\frac{\rho^0\left[\overline{\mathbf{J}}-\langle \mathbf{J}^k\rangle-\mathbb{H}^0\ast \left(\frac{1}{\sigma} \mathbf{J}^k\right)\right]-\mathbb{G}^0\ast \mathbf{J}^k}{1/\sigma+\rho^0}.
\end{equation}

All of these methods involve a reference conductivity $\sigma^0$, or a reference resistivity $\rho^0$. Whereas the final result is in principle independent of these quantities, their {\red values [instead of `value']} have a dramatic influence on the convergence properties of the algorithms. Notably, optimal convergence of the 'accelerated' scheme is obtained with the choice~\cite{Eyre99}
\begin{equation}
\label{eq:opteyre}
\sigma^0=-\sqrt{\sigma_1\sigma_2},
\end{equation}
where the use of a negative reference conductivity (devoid of physical meaning) is warranted by the arbitrary character of the reference medium. In this connection, we point out that in Ref.~\cite{moulinec13}, which addresses the analogous elasticity problem, the reference stiffness moduli have their sign changed, which avoids dealing with negative values.

For the ``direct" scheme, optimal convergence properties were studied in the context of elasticity~\cite{moulinec1998numerical}. Adapting the method used in the latter reference to the conductivity problem, it is straightforward to show that the corresponding optimal choice is
\begin{equation}
\label{eq:optdirect}
\sigma^0\approx \frac{1}{2}(\sigma_1+\sigma_2),
\end{equation}
a result to be used extensively below.

\section{Classical and modified Green operators}\label{sec:green}
In practice, the domain $\Omega$ is discretized as a two-dimensional (2D) pixel image, or three-dimensional (3D) voxel image. The convolution product $G^0_{ij}\ast P_j$ in~(\ref{eq:ls1}) is evaluated in the Fourier domain as
\begin{equation}\label{eq:greenFour}
\int_\Omega {\rm d^d}\,\mathbf{x'}G^0_{ij}(\mathbf{x}-\mathbf{x'})P_j(\mathbf{x'}) \approx
  \frac{1}{L^d}\sum_\mathbf{q} G^0_{ij}(\mathbf{q})P_j(\mathbf{q})\ee^{\ii\mathbf{q}\cdot\mathbf{x}},
\end{equation}
where the Fourier mode components take on values $q_i=(2\pi/L)(-L/2,...,L/2-1)$ ($i=1$, ..., $d$), and $L$ is measured in pixel/voxel size units. The vector $P_j(\mathbf{q})$ is the Fourier transform
\begin{equation}
P_j(\mathbf{q})=\sum_\mathbf{x} P_j(x)\ee^{-\ii\mathbf{q}\cdot\mathbf{x}},
\end{equation}
where the sum is over all pixels/voxels $\mathbf{x}$ in $\Omega$. Classically, the Fourier transform of the Green operator used in (\ref{eq:greenFour}) is approximated by its continuum expression
\begin{equation}\label{eq:greenCont}
G^0_{ij}(\mathbf{q})=\int {\rm d}^d\!\mathbf{x}\,G^0_{ij}(\mathbf{x})\ee^{-\ii\mathbf{q}\cdot\mathbf{x}}=\frac{q_iq_j}{\sigma^0|q|^2},
\end{equation}
where the integration is over the infinite domain and $|q|=\sqrt{q_kq_k}$. We call hereafter this version of the Green operator the `continuous' Green operator. This name is choosen as a matter of convenience as the operator $\mathbb{G}^0$ is only the discretization, on a regular grid, in the Fourier domain, of the continuum Green operator.

\begin{figure}
\centering
    \includegraphics[width=6cm]{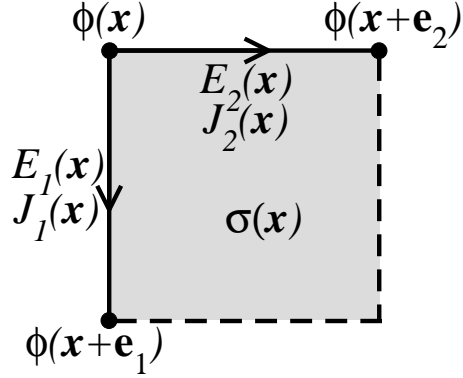}
\caption{
  \label{fig:schema}
  2D pixel at point $\mathbf{x}$ with superimposed resistor network (see Equation~\ref{eq:rn});
  here $\mathbf{e}_1$ is oriented from top to bottom and $\mathbf{e}_2$ left to right.
  }
\end{figure}

On the other hand, intrinsically discrete schemes can be considered. For instance, in the context of continuum mechanics, modified Green operators have been introduced, where partial derivatives are approximated by centered~\cite{Muller96} or forward~\cite{willot08c} differences. In the conductivity problem, the latter discretization amounts to solving a resistor network problem~\cite{luck1991conductivity}
\begin{equation}\label{eq:rn}
\partial_i J_i(\mathbf{x})\approx J_i(\mathbf{x})-J_i(\mathbf{x}-\mathbf{e}_i), \qquad
\partial_i \Phi(\mathbf{x})\approx \Phi(\mathbf{x}+\mathbf{e}_i)-\Phi(\mathbf{x}),
\end{equation}
where  $J_i(\mathbf{x})$ represents the current along the bond pointing in the direction $\mathbf{e}_i$ from point $\mathbf{x}$, and $\Phi(\mathbf{x})$ is the potential at node $\mathbf{x}$.
The same fields are used as approximations of the exact solution in a continuous medium.
The nodes in the network are mapped to the corners of each voxel and the bonds are mapped to the edges (see Figure~\ref{fig:schema}). In this setting, the electric field and current are estimated at edge centers, which turns (\ref{eq:rn}) into the centered scheme
\begin{equation}\label{eq:rn2}
\partial_i J_i(\mathbf{x})\approx J_i\left(\mathbf{x}+\frac{\mathbf{e}_i}{2}\right)-J_i\left(\mathbf{x}-\frac{\mathbf{e}_i}{2}\right), \qquad
-E_i\left(\mathbf{x}+\frac{\mathbf{e}_i}{2}\right)=\partial_i\Phi\left(\mathbf{x}+\frac{\mathbf{e}_i}{2}\right)\approx \Phi(\mathbf{x}+\mathbf{e}_i)-\Phi(\mathbf{x}).
\end{equation}
Here again, derivatives are approximated by differences over points separated by one voxel size, unlike in~\cite{Muller96}. Discretizations~(\ref{eq:rn2}) and~(\ref{eq:rn}) are equivalent up to a translation of $J_i$ and $E_i$ by a vector $\mathbf{e}_i/2$, provided that $\sigma$ is constant in each voxel (see Figure~\ref{fig:schema}). For simplicity, we use~(\ref{eq:rn}) hereafter. The `discrete' Green operator $\widetilde{\mathbb{G}^0}$ entering the corresponding Lippmann-Schwinger equation reads~\cite{luck1991conductivity,willot08c}
\begin{equation}\label{eq:greendisc}
 \widetilde{G}^0_{ij}(\mathbf{k})=\frac{k_ik_j^*}{\sigma^0|k|^2}, \qquad k_i=\ee^{\ii q_i}-1=2\ii\sin(q_i/2)\ee^{\ii q_i/2},
\end{equation}
where $|k|=\sqrt{k_ik_i^*}$ and $^*$ is the complex conjugate. In the Fourier domain, the `discrete' gradient, divergence and Laplacian operators amount to multiplications by $k_i$, $-k_i^*$ and $|k|^2$, respectively, instead of $\ii q_i$, $\ii q_i$ and $|q|^2$ when using the continuum Green operator~$\mathbb{G}^0$. Likewise, the terms `divergence-free' and `compatible' depend on the employed discretization. In the long-wavelength limit $\mathbf{q}\to 0$, these differences disappear and equation~(\ref{eq:greendisc}) reduces to (\ref{eq:greenCont}). In the dual setting, the discrete Green operator associated to the current is defined, mutatis mutandis, as in Equation~(\ref{eq:greendual}). Hereafter, the operator $\widetilde{\mathbb{G}^0}$ is referred to as the `discrete' Green operator.

The representation of the problem in terms of a resistor network result in several useful properties. First, contrary to the variational algorithm~\cite{Brisard10}, the solution
does not depend on the choice for the reference material $\sigma^0$. Second, the operator $\widetilde{\mathbb{G}^0}$ is a smooth periodic function, where contrary to $\mathbb{G}^0$, high-frequencies are cut out in the Fourier domain. This is expected to result in better convergence properties. Third, the discretization in (\ref{eq:rn}) enforces local current conservation, which makes Kirchhoff's law hold at each node. Consequently, the outward flow of $\mathbf{J}$ along a closed surface, defined as a sum of currents over the bonds that pierce the surface, is zero.

As long as they converge, all numerical schemes must deliver the same results for a given choice of Green operator. Conversely, choosing one Green operator will select one particular approximation to the solution of the problem considered. It is the purpose of this work to assess the advantages, from the numerical viewpoint, in the context of electrical conductivity, of using $\widetilde{\mathbb{G}^0}$ in place of $\mathbb{G}^0$.

In this paper, the direct (DS), accelerated (AS), augmented-Lagrangian (AL), and variational (VAR) schemes are investigated. We also consider the dual versions of DS and AL, denoted by \X{DS} and \X{AL}, respectively. All of these make use of the continuous Green operator $\mathbb{G}^0$. Same algorithms, but with the \emph{discrete} Green operator $\widetilde{\mathbb{G}}^0$ instead of $\mathbb{G}^0$ are also examined. They are referred to with a `tilde' notation as \D{DS}, \D{AS}, \D{AL}, \D{VAR}, \DX{DS}, and \DX{AL}.
We emphasize that the results presented here for the variational approaches VAR and \D{VAR} make use of the Green operators $\mathbb{G}^0$ and $\widetilde{\mathbb{G}^0}$ rather than of the more complex discretization proposed in~\cite{Brisard10}. Also, in the latter approaches, definite-positiveness of matrix $\mathcal{M}$ (see Sec.\ \ref{sec:FFTmethods}) is not guaranteed in the conjugate-gradient procedure. This specific issue has not been considered further as numerical experiments that we performed indicate that the latter schemes nevertheless converge.

% -------------------------- LOCAL FIELDS-----------------------------------
\section{A stiff case: fields in the four-cell microstructure}
\label{sec:local}
\begin{figure}[!ht]
  \begin{center}
    \includegraphics[width=8cm]{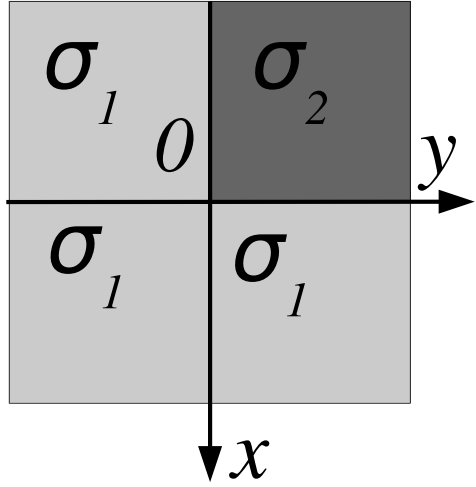}
    \caption{\label{fig:4cells}
    Elementary periodic domain $\Omega=(-L/2,+L/2)^2$ with four-cell microstructure. The inclusion has conductivity $\sigma_2$ and the matrix has conductivity $\sigma_1$.
}
  \end{center}
\end{figure}

The `four-cell' microstructure is one of the few periodic structures for which an exact solution~\cite{craster01} is available. We consider the special case, represented in Figure~\ref{fig:4cells}, where the elementary cell is made of a single square inclusion of surface fraction $25$\%. Because of the presence of corners, fields are singular in the infinite-contrast limit, which makes this case a good benchmark for numerical methods. In this Section, numerical results for the current computed with either the continuous Green operator $\mathbb{G}^0$ or the discrete operator $\widetilde{\mathbb{G}}^0$ are compared with the exact solution. The inclusion is highly conducting, with a contrast ratio $\sigma_2/\sigma_1=2\times 10^3$.
% ----------------------------------------
% ----------------------------------------

\begin{figure}[!ht]
  \begin{center}
   \begin{tabular}{cccc}
     \includegraphics[width=3cm]{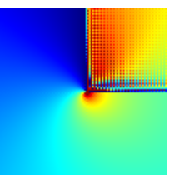} &
     \includegraphics[width=3cm]{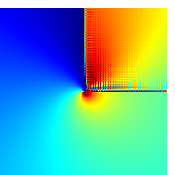} &
     \includegraphics[width=3cm]{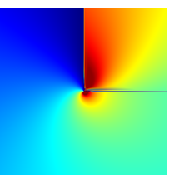} &
     \includegraphics[width=3cm]{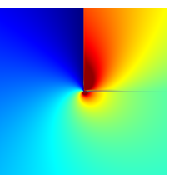} \\
     \includegraphics[width=3cm]{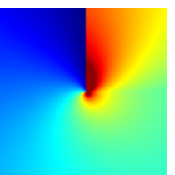} &
     \includegraphics[width=3cm]{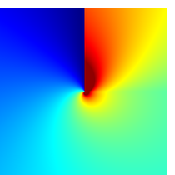} &
     \includegraphics[width=3cm]{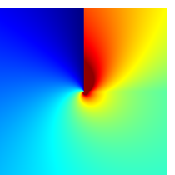} &
     \includegraphics[width=3cm]{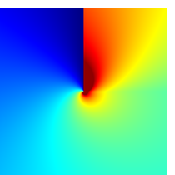} \\
     $L=1024$ & $L=2048$ & $L=4096$ & $L=8192$
    \end{tabular}
    \caption{\label{fig:localFieldMaps}
    Four-cell microstructure of Figure~\ref{fig:4cells}. Maps of the vertical current component $J_1(x,y)$ in the region $-0.05 L\leq x, y\leq 0.05 L$, for increasing resolution $L$ (as indicated). Top: with continuous Green operator $\mathbb{G}^0$. Bottom: with discrete operator $\widetilde{\mathbb{G}}^0$.
    }
  \end{center}
\end{figure}

The behavior of the electric current near the singular corner at point $(x,y)=(0,0)$ is illustrated in Figure~\ref{fig:localFieldMaps}. Maps of the vertical component $J_1(x,y)$ obtained with $\mathbb{G}^0$ (top) and $\widetilde{\mathbb{G}}^0$ (bottom) are displayed for increasing resolutions (left to right). Only the small region $-5.10^{-2}L\leq x, y\leq 5.10^{-2}L$ around the corner is shown.  Numerical artifacts in the highly-conducting phase are conspicuous when using the continuous Green operator $\mathbb{G}^0$. They consist of high-frequency oscillations all over the conducting region, particularly near the horizontal interface~\cite{note0}, where the represented field component should be continuous. Such oscillations are almost absent when using $\widetilde{\mathbb{G}}^0$.
% ----------------------------------------
% ----------------------------------------

\begin{figure}[!ht]
  \begin{center}
    \includegraphics[width=12cm]{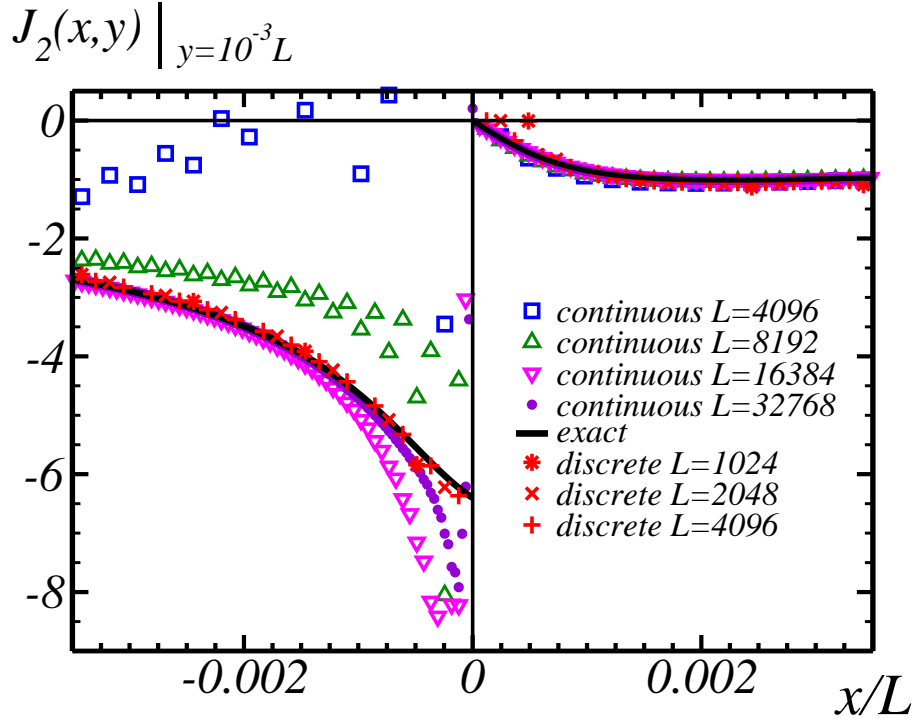}
    \caption{\label{fig:localFieldplots}
    Four-cell microstructure of Figure~\ref{fig:4cells}. Values of the horizontal current component $J_2(x,y=10^{-3}L)$ vs.\ $x$, for various resolutions $L$ (as indicated). Solid black: exact solution. Markers $\ast$,  $\times$ and  $+$ (red): FFT results with discrete Green operator $\widetilde{\mathbb{G}}^0$. Other markers and colors: FFT results with continuous Green operator $\mathbb{G}^0$.}
  \end{center}
\end{figure}

Figure~\ref{fig:localFieldplots} displays plots of the horizontal component $J_2(x,y)$ versus $x$ at $y=10^{-3}L$, close to the inclusion boundary. Negative values of $x$ correspond to the interior of the inclusion. Numerical results computed with both Green operators are compared with the exact solution. To draw meaningful graphs, data points obtained with $\mathbb{G}^0$ were post-processed prior to plotting by convolution over a window of $2\times 2$ adjacent pixels. This crude filtering device greatly reduces oscillations. Results obtained with $\widetilde{\mathbb{G}}^0$ have not been modified.
Given sufficient resolution all methods converge to the exact solution. However, although all methods lead to almost identical solutions in the matrix, results strongly differ in the highly conducting region. The figure, which represents calculations carried out for various resolutions, shows that employing $\widetilde{\mathbb{G}}^0$ makes convergence notably easier. Indeed, data points obtained with $\widetilde{\mathbb{G}}^0$ at moderate resolution $L=1024$ are much closer to the exact solution than those obtained from $\mathbb{G}^0$ at the highest resolution $L=32\,768$.
% ----------------------------------------
% ----------------------------------------

In a previous study involving porous media~\cite{willot08c}, the continuous Green operator was already observed to induce awkward aliasing effects at high contrast. They usually take place near interfaces involving a region where the field considered is not uniquely defined in the infinite-contrast limit (e.g., the strain in a pore, or the electric current in an infinitely conducting inclusion).

\section{Convergence rate}\label{sec:results}
This Section further examines for a few selected microstructures the convergence properties of FFT schemes. Algorithmic convergence being harder in the case of strongly contrasted composites, the quantity of interest here is the number of iterations as a function of the contrast ratio $\sigma_2/\sigma_1$.

\subsection{Convergence criteria}\label{subsec:criteria}
Convergence criteria can be written either in the direct or Fourier representations. The most compelling ones are those that include high Fourier frequency behavior~\cite{moulinec13}. In relation to FFT algorithms, the following criteria are considered:
\begin{subequations}
\label{eq:precision}
\begin{eqnarray}
\label{crit:divergence}
  \eta_1 &=&
      \|\overline{\mathbf{J}}\|^{-1} \max_\mathbf{x} \left| \textnormal{FT}^{-1}\left\lbrace k_i^*(\mathbf{q})J_i(\mathbf{q}); \mathbf{x}\right\rbrace\right|\leq \epsilon,\\
  \label{crit:equilibrium}
  \eta_2 &=&
      \|\overline{\mathbf{E}}\|^{-1} \max_{i\neq j, \mathbf{x}} \left| \textnormal{FT}^{-1}\left\lbrace k_i(\mathbf{q})E_j(\mathbf{q})-k_j(\mathbf{q})E_i(\mathbf{q}); \mathbf{x}\right\rbrace\right|\leq \epsilon,
\end{eqnarray}
\end{subequations}
where $\epsilon\ll 1$ is the required precision and $\textnormal{FT}^{-1}$ is the backward Fourier transform. Criterion (\ref{crit:divergence}) puts emphasis on the current conservation, whereas (\ref{crit:equilibrium}) imposes compatibility; apart from a difference in the norm used, they are akin to those used in~\cite{moulinec13}. These equations refer to the discrete Green operator $\widetilde{\mathbb{G}^0}$.
Current conservation and compatibility are enforced differently when using the continuous Green operator $\mathbb{G}^0$. In the latter case, $\mathbf{k}(\mathbf{q})$ and $\mathbf{k}^*(\mathbf{q})$ are replaced by $\mathbf{q}$ in Equation~(\ref{eq:precision}).

Among the computational schemes introduced in Section 4, DS and \D{DS} enforce compatibility, at each iteration, which trivially guarantees that $\eta_2 = 0$. Instead, electric current conservation in the form of the equality $\eta_1 = 0$ is enforced by the dual schemes \X{DS} and \DX{DS}. On the other hand, the remaining schemes in general lead to nonzero values of $\eta_1$ and $\eta_2$. This suggests using as a convergence criterion the inequality $\eta\leq\epsilon$ where $\eta=\eta_1$ for the primary (non-dual) schemes DS, \D{DS}, AL, \D{AL}, AS, \D{AS}, VAR, and where $\eta=\eta_2$ for the dual ones \X{DS}, \DX{DS}, \X{AL}, \DX{AL}.

\subsection{Test microstructures}
\begin{figure}
\begin{center}
\includegraphics[width=13cm]{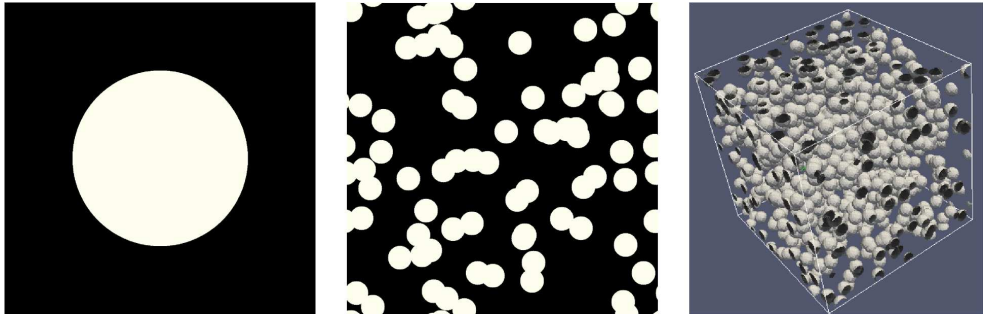}
\end{center}
\caption{\label{fig:micro} Elementary cell $\Omega$ of the ``2D-periodic'' microstructure (left), and the 2D (center) and 3D (right) random periodic Boolean models. Surface and volume fractions of the inclusions are, respectively, $25$, $30$ and $20\%$.}
\end{figure}

Convergence rates are monitored for three microstructures, periodic in all directions, whose unit cells $\Omega$ are represented in Figure~\ref{fig:micro}. The leftmost 2D cell, of size $L=1024$ pixels, contains a single circular disk-shaped inclusion of surface fraction $25\%$. This system is simply referred to as the `2D-periodic' medium hereafter. The middle cell is a random 2D Boolean model of size $L=1024$ built from disks of diameter 80 pixels, of overall surface fraction $30\%$. The rightmost cell is a random 3D Boolean model of size $L=256$, made of spherical inclusions of diameter 20 voxels, with overall volume fraction $20\%$.
% ----------------------------------------
% ----------------------------------------

\subsection{2D periodic medium}
\begin{figure}[!ht]
\begin{center}
\includegraphics[width=10cm]{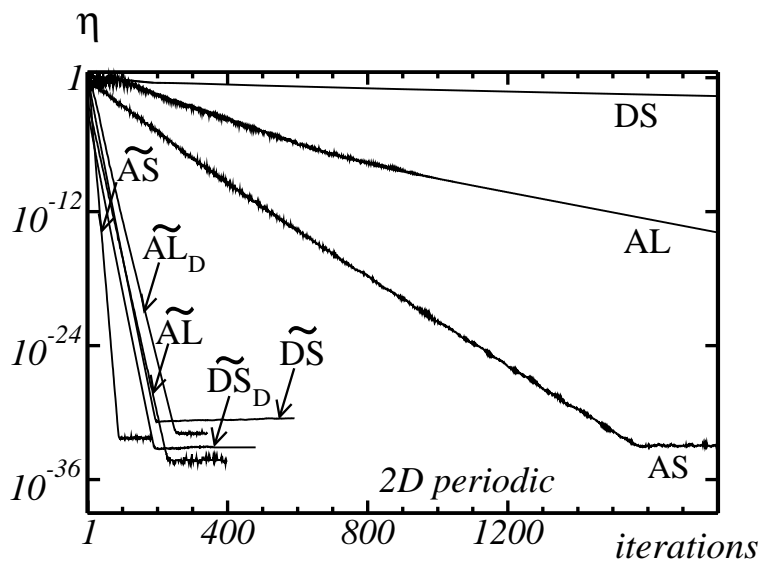}
\caption{\label{fig:niter0} ``2D-periodic" medium. Convergence indicator $\eta$ vs.\ number of iterations in logarithmic-linear scale, for various FFT schemes: using the continuous Green operator (DS, AS, and AL), and the discrete Green operator (\D{DS}, \DX{DS}, \D{AS}, \D{AL} and \DX{AL}).}
\end{center}
\end{figure}

Figure (\ref{fig:niter0}) illustrates for some of the algorithms introduced in Section 4 applied to the `2D-periodic' medium how the indicator $\eta$ tends to zero as the number of iterations increases. The contrast ratio is fixed at $\sigma_2/\sigma_1=2\times 10^3$. For exploratory purposes, quadruple precision was used in these calculations to allow for tiny values of $\eta$. Prior to drawing the plots, the quantities $\sigma^0$ and $\rho^0$ were optimized manually to minimize the number of iterations needed to reach the arbitrary threshold $\eta<\epsilon=10^{-12}$. For all methods, $\eta$ decreases exponentially with the number of iterations down to some constant value determined by machine precision. Roughly, algorithms separate in two classes. The first one comprises the continuous schemes, namely, DS, AL and AS, which are the slowest converging ones. However, in this class and for the microstructure considered, Eyre and Milton's AS is clearly superior. The simple DS is by far the worst, and the AL scheme is intermediate. The other class encompasses the `discrete' schemes (primary and dual). They all make $\eta$ saturate in less than 300 iterations, which is another hint at the good behavior of the discrete Green operator. In that class, Eyre and Milton's method (\D{AS}) again proves the fastest converging one.
% ----------------------------------------
% ----------------------------------------

The optimal reference conductivity $\sigma^0$ and resistivity $\rho^0$ used in Figure~\ref{fig:niter0} are summarized in the second column of Table~\ref{tab:ref}. The integer number in brackets is the number of iterations needed to reach the threshold $\eta<\epsilon=10^{-8}$, which in practice is a good trade-off between speed and accuracy. As already mentionned, Equation~(\ref{eq:optdirect}) optimizes the DS with the continuous Green operator. It gives $\sigma^0=1000.5$ and ---this is an empirical finding--- also optimizes \D{DS} with the discrete Green operator. Introducing  phase resistivities as $\rho_{1,2}=1/\sigma_{1,2}$, an analogous formula (easy to demonstrate in the continuum) holds for the optimal resistivity in the continuous dual `direct' scheme \X{DS}, namely,
\begin{equation}
\rho^0=\frac{1}{2}(\rho_1+\rho_2),
\end{equation}
which gives here $\rho^0\simeq 0.5$. Again empirically, we find that this value optimizes  as well the discrete dual `direct' scheme \DX{DS}. As expected, the optimum $\sigma^0\simeq -44.7$ reported for AS matches Eyre and Milton's result, Equation~(\ref{eq:opteyre}). However, although negative, the optimum $\sigma^0$ found for \D{AS} is \emph{not} consistent with this formula. Finally, the values reported for the primary augmented-Lagrangian schemes AL and \D{AL} and their dual versions do not match any of the previous analytical estimates.

% ----------------------------------------
\newcommand{\zz}{\vphantom{\resizebox{1.2cm}{!}{Foo}}}
\begin{table}
\begin{center}
\begin{tabular}{lccc}\toprule
 \multirow{2}{*}{} & \multicolumn{3}{c}{$\sigma^0$} \\
 \cmidrule{2-4} & ``2D-periodic" & 2D Boolean & 3D Boolean \\ \midrule
\zz{DS}      & $1000.5$ ($15621$)  & $(\sigma_1+\sigma_2)/2$                                  & $(\sigma_1+\sigma_2)/2$            \\ \hline
\zz{AL}      & $76$ ($1556$)       & $3.\,10^{-3}\sigma_1+1.8\sqrt{\sigma_1\sigma_2}$         & $1.7\sqrt{\sigma_1\sigma_2}$       \\ \hline
\zz{AS}      & $-44.7$ ($663$)     & $-\sqrt{\sigma_1\sigma_2}$                               & $-\sqrt{\sigma_1\sigma_2}$         \\ \hline
\zz{VAR}     & N/A                 & $0.50(\sigma_1+\sigma_2)$                                & N/A                                 \\ \hline
\zz{\D{DS}}  & $1000.5$ ($46$)     & $0.50(\sigma_1+\sigma_2)$                                & $0.53\sigma_1+0.50\sigma_2$        \\ \hline
\zz{\D{AL}}  & $1855$ ($95$)       & $0.30(\sigma_1+\sigma_2)$                                & $0.56\sigma_1+0.26\sigma_2$        \\ \hline
\zz{\D{AS}}  & $-1390$ ($46$)      & $-0.30(\sigma_1+\sigma_2)$                               & \ytab{-(1/3.6)\sigma_1}{-3.6\sigma_1} \\ \hline
\zz{\D{VAR}} & N/A                 & $0.50(\sigma_1+\sigma_2)$                                & N/A                                 \\
 \midrule  & \multicolumn{3}{c}{$\rho^0$} \\  \cmidrule{2-4}
\zz{\X{DS}}  & $0.5$ ($14616$)     & $(\rho_1+\rho_2)/2$                                      & $(\rho_1+\rho_2)/2$                \\ \hline
\zz{\X{AL}}  & $0.033$ ($1336$)    & $3\,10^{-3}\rho_1+1.8\sqrt{\rho_1\rho_2}$                & $1.7\sqrt{\rho_1\rho_2}$           \\ \hline
\zz{\DX{DS}} & $0.5$ ($46$)        & $0.50(\rho_1+\rho_2)$                                    & $0.48\rho_1+0.52\rho_2$            \\ \hline
\zz{\DX{AL}} & $1.09$ ($93$)       & $0.30(\rho_1+\rho_2)$                                    & $0.40\rho_1+0.55\rho_2$            \\
 \bottomrule \end{tabular}
\vspace{0.5cm}
\caption{\label{tab:ref}
Optimal reference conductivities $\sigma^0$ and resistivities $\rho^0$ determined for the indicated FFT schemes. Values given for the ``2D-periodic" microstructure correspond to the contrast ratio $\sigma_2/\sigma_1=2\times 10^3$, with the number of iterations indicated in brackets. For Boolean models, the formulas given are consistent with the behavior observed at high contrast, although the low-contrast behavior
may slightly differ. Those for schemes DS, AS and \X{DS} are exact ones. Missing entries (N/A) indicate that the corresponding schemes have not been investigated.}
\end{center}
\end{table}
% ----------------------------------------

\subsection{2D and 3D Boolean media: reference conductivity or resistivity}
\label{subsec:2drandom}
A more thorough study was carried out for the Boolean models, in which the optimal reference conductivity $\sigma^0$ or resistivity $\rho^0$ was measured as a function of the contrast.

In order to avoid unnecessary long computations, the reference was first manually optimized on a low-resolution grid of size $L=64$ (in 2D) or $L=32$ (in 3D). The optimized reference was then tested on a full-resolution grid of size $L=1024$ (2D) or $L=256$ (3D). In all but a few cases, the number of iterations to convergence found with the low-resolution and high-resolution grids was nearly the same. The number of iterations found on the full-resolution grid was kept if the difference was less than $10\%$; otherwise, the reference was optimized again, this time on the full-resolution grid, to provide a definitive number of iterations. Manual optimization of the reference parameters was carried out following a rough dichotomy procedure, disregarding for simplicity the possibility of concurrent local optima. The convergence criterion was set to $\eta\leq \epsilon=10^{-8}$ in these calculations.

Our findings are summarized in the third and fourth columns of Table~\ref{tab:ref}, where the formulas given essentially represent high-contrast behaviors in the regimes $\sigma_2/\sigma_1\ll 1$ or $\sigma_2/\sigma_1\gg 1$. Indeed, in some cases, the  low-contrast behavior may differ from that given (see succeeding text).

At the exception of scheme \D{AS} in the 3D Boolean medium, for which $\sigma^0/\sigma_1$ tends to a constant at high contrast ---notice the symmetry between both high-contrast regimes, the behaviors we observed are of the following types:
\begin{subequations}
\begin{align}
\label{eq:lin}
\sigma^0/\sigma_1&=\alpha_1+\alpha_2\,r,\\
\label{eq:sqrt}
\text{or }\quad\sigma^0/\sigma_1&=\beta_1+\beta_2\,r^{1/2},
\end{align}
\end{subequations}
where $r=\sigma_2/\sigma_1$, and $\alpha_{1,2}$ and $\beta_{1,2}$ are numerical constants of various signs (see Table I). These forms generalize Equations (\ref{eq:opteyre}) and (\ref{eq:optdirect}). They apply to the `primary' schemes, and similar ones hold for the `dual' schemes with $\sigma$ substituted by $\rho$. When nonzero, the coefficient $\beta_1$, of order $10^{-3}$, is of unclear origin. The coefficients reported in the table were determined by nonlinear least-square fitting on our data. Additional fitting attempts with functional forms other than (but related to) those retained indicate that the first digit of the coefficients is significative, whereas the error on the second one is hard to evaluate. Different coefficients $\alpha_1$ and $\alpha_2$ are provided when our results do not support an equality $\alpha_1=\alpha_2$. However, our results strongly suggest that $\alpha_1=\alpha_2$ for the 2D Boolean system whenever Equation~(\ref{eq:lin}) applies, while this symmetry does not carry over to the 3D case, except for the DS, where $\alpha_1=\alpha_2=1/2$ (exact) in two and three dimensions.

Although the optimum may in some cases be of the same form with the continuous and discrete Green operators, there are other cases such as with AS and \D{AS}, for which the optimal forms look strongly dissimilar. Moreover, comparing columns 2 and 3 of the table for the contrast $\sigma_2/\sigma_1=2\times 10^3$ indicates that the optima found somewhat depend on the microstructure.

\begin{figure}[!ht]
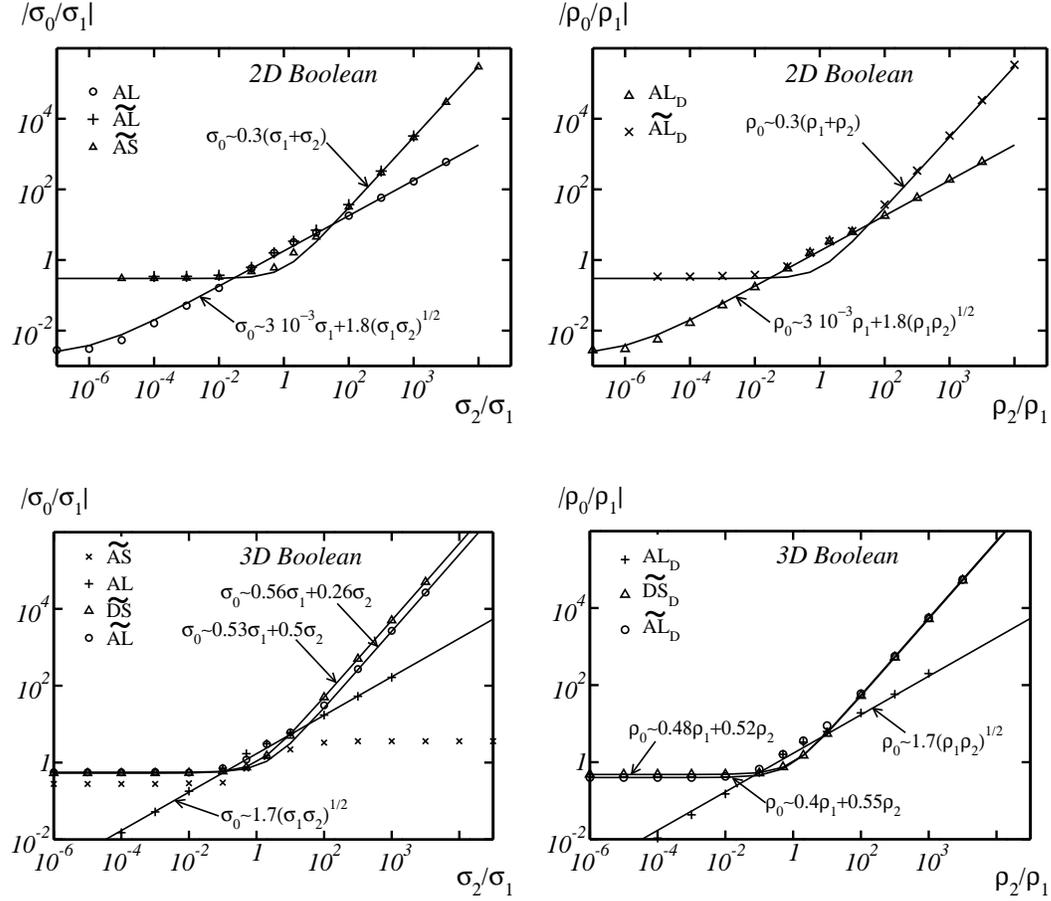

\begin{center}
 \includegraphics[width=6.7cm]{fig6a.eps}\quad
 \includegraphics[width=6.7cm]{fig6b.eps}
 \medskip

 \includegraphics[width=6.7cm]{fig6c.eps}\quad
 \includegraphics[width=6.7cm]{fig6d.eps}
\caption{\label{fig:fits} 2D (top) and 3D (bottom) Boolean models. Absolute value of the normalized optimal conductivity $|\sigma^0/\sigma_1|$ vs.\ $\sigma_2/\sigma_1$ (left), and optimal resistivity $|\rho_0/\rho_1|$  vs.\ $\rho_2/\rho_1$ (right), for the schemes indicated. Symbols: FFT results. Solid: numerical fits (see Table~\ref{tab:ref}).}
\end{center}
\end{figure}

The behaviors gathered in the table are supported by Figure~\ref{fig:fits}, which presents plots of our 2D and 3D data and the corresponding fitting curves. The `primary' and `dual' schemes are addressed in separate plots. The signs indicated in the Table cannot be read from the figures, where absolute values are displayed in logarithmic scale. In the 2D Boolean model, the data for the primary schemes and for their dual are numerically quite close in this mode of representation, so that the left and right plots superimpose almost exactly. Interestingly, the plots reveal the unique non-trivial behavior of the discrete schemes \D{AL} and \DX{AL} in the low-contrast region $0.1\leq\sigma_2/\sigma_1\leq 10$, where they behave as $\sqrt{\sigma_2/\sigma_1}$ even though the linear behavior reported in Table~\ref{tab:ref} takes place at higher contrasts. On the other hand, the continuous schemes AL and {\red\X{AL} $[$ instead of `${\rm AL^D}$' $]$} essentially behave as a square root for all contrasts (up to a small corrective term in 2D cases). As already noticed in the discussion of the table, the discrete 3D `accelerated' scheme \D{AS} with its intriguing asymptotic behavior (constant on both sides of the contrast range) stands as an outlier. For it no fit has been attempted. We emphasize that in all cases examined with the `accelerated' schemes, the optimal square-root estimate (\ref{eq:opteyre}) ---exact in scheme AS--- yields poor convergence when applied to \D{AS}.
% ----------------------------------------
% ----------------------------------------

In 2D, the formula $\sigma_0=0.5(\sigma_1+\sigma_2)$ indifferently optimizes the discrete and continuous \D{VAR} and VAR schemes. We observed similar convergence rates, up to $3$\% difference in the number of iterations, for these algorithms within the range $0.4\leq \sigma_0/(\sigma_1+\sigma_2)\leq 0.9$. However, outside of this range, the convergence of the VAR scheme deteriorates. The small sensitivity with respect to the reference material $\sigma_0$ in this method is supported by other studies~\cite{Gelebart}.

\begin{figure}[!ht]
\begin{center}
\vspace{0.5cm}
\includegraphics[width=10cm]{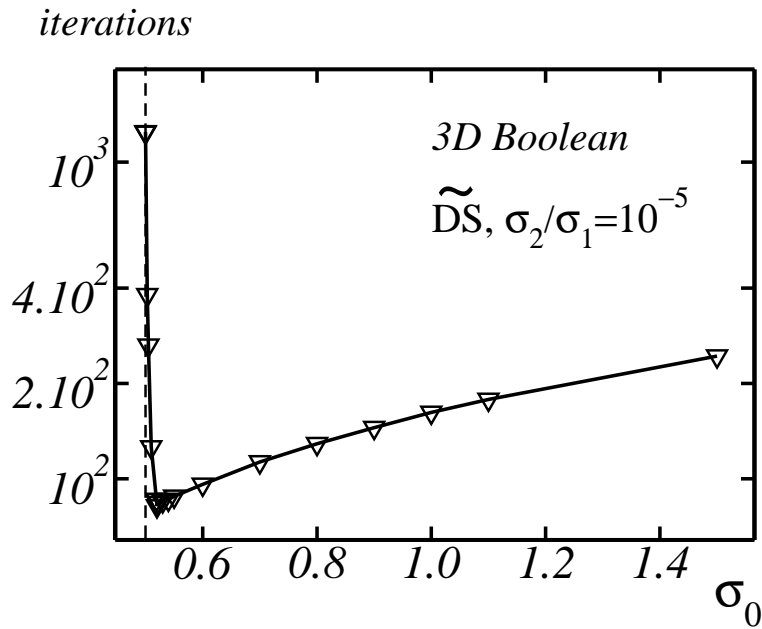}
\end{center}
\caption{\label{fig:convRef} 3D Boolean model. Number of iterations vs.\ reference conductivity $\sigma^0$ for the direct scheme with discrete Green operator (\D{DS}). The contrast ratio is $\sigma_2/\sigma_1=10^{-5}$. The convergence criterion is $\eta\leq \epsilon=10^{-8}$. The value $\sigma^0=(\sigma_1+\sigma_2)/2$ is represented by the vertical dotted line. The solid line between data points is a guide to the eye.
}
\end{figure}

We also investigated the sensitivity to the choice of $\sigma^0$ in the `direct' discrete schemes. In the 2D Boolean model and for the discrete scheme \D{DS}, the choice $\sigma^0=0.50(\sigma_1+\sigma_2)$ proves optimal, which matches the exact result relative to DS. However, with \D{DS}, nearly optimal 2D results are also obtained with choices $\sigma^0<(\sigma_1+\sigma_2)/2$. By contrast, in 3D, the number of iterations may be extremely sensitive to the choice of $\sigma^0$. Figure~\ref{fig:convRef} illustrates this. It represents the number of iterations versus $\sigma^0$ for \D{DS} in the 3D Boolean model, with contrast $\sigma_2/\sigma_1=10^{-5}$. No convergence is observed for $\sigma^0<0.5(\sigma_1+\sigma_2)$, and the optimal choice is about $\sigma^0\approx 0.53\,\sigma_1$.
% ----------------------------------------
% ----------------------------------------

\subsection{2D and 3D Boolean media: convergence properties}
\begin{figure}[!ht]
\begin{center}
 \begin{tabular}{c}
   \includegraphics[width=11cm]{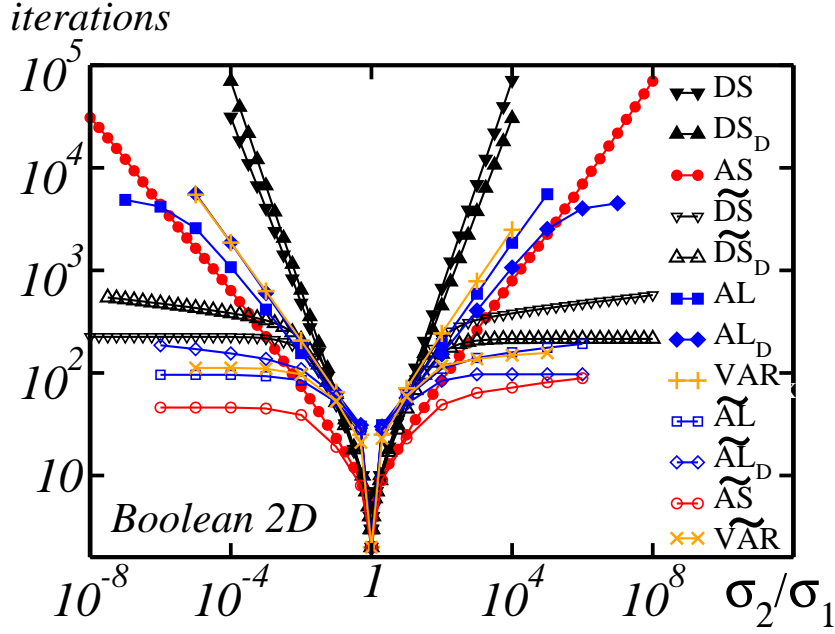}\\ \\
   \includegraphics[width=11cm]{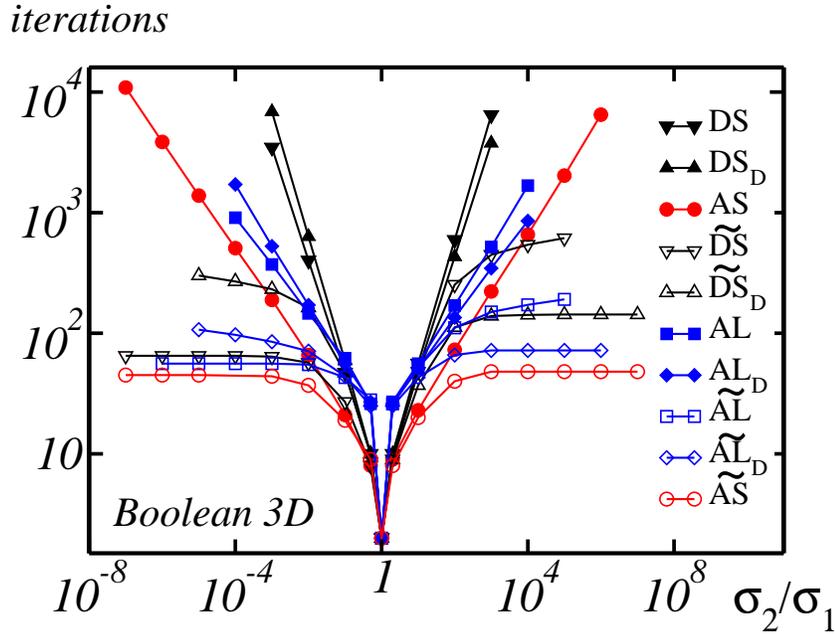}
 \end{tabular}
\caption{\label{fig:conv} 2D and 3D Boolean models (top and bottom). Number of iterations vs.\ contrast for various FFT algorithms. The convergence criterion is $\eta\leq \epsilon=10^{-8}$. Solid lines between data points are guides to the eye.}
\end{center}
\end{figure}

This Section examines convergence performance for the 2D and 3D Boolean models, expressed by the number of iterations $N$ as a function of the contrast ratio $r=\sigma_2/\sigma_1$. Figure (\ref{fig:conv}) illustrates the performance of the various FFT schemes considered, in calculations optimized by using the reference conductivity or resistivity discussed in the previous section. Schemes using $\mathbb{G}^0$ are represented by filled symbols and the  $+$ marker, whereas discrete schemes using $\widetilde{\mathbb{G}}^0$ are represented by empty symbols and the $\times$ marker.
% ----------------------------------------
% ----------------------------------------

We recover known results of linear scaling $N\sim r$ for DS and \X{DS}, and of square-root scaling $N\sim r^{1/2}$ for AS~\cite{Eyre99}. Similar convergence rates are observed for AL and \X{AL}, and {\red for VAR [instead of `for the VAR']}. As a rule, given the FFT method, the `primary' scheme always converges better than the `dual' one when $r<1$, while the opposite holds when $r>1$. For instance, at very strong contrast ratio $r>10^7$, the convergence of the dual `augmented-Lagrangian' scheme \X{AL} is much faster than that of the primary one AL.

As to discrete schemes, they are much more efficient than their continuous counterparts. For discrete schemes, $N(r)$ is either a bounded or slowly increasing function of $r$, which shows that using the discrete Green operator $\widetilde{\mathbb{G}}^0$ definitely provides a dramatic improvement of convergence. By optimizing the choice between the `primary' or `dual' versions of the discrete algorithm at hand depending on whether $r<1$ or $r>1$, one can even achieve \emph{convergence in a finite number of iterations} in the infinite-contrast limit.

Overall, the figure shows that among all schemes the discrete version \D{AS} of the AS is the better converging one in 2D and 3D.

\section{Optimizing the ``direct" scheme with discrete Green operator}
\label{sec:speed}
In applications dealing with large microstructures (typically, multiscale materials) fast and memory-efficient implementations of FFT methods are required. One common way of minimizing both CPU speed and memory storage is to recompute the Green operator at each iteration. As long as the Green operator is easy to compute, this strategy is usually faster than storing a very large tensor field. This is used in the CraFT~\cite{craft} and morph-Hom~\cite{morphhom} softwares. As an example, a low-cost implementation of DS is as follows:

Initialization: set $A_i(\mathbf{x})\equiv 0$.
\begin{enumerate}
  \item Set $A_i(\mathbf{x}):=[\sigma(\mathbf{x})-\sigma^0]A_i(\mathbf{x})$;
  \item Set $A_i(\mathbf{q}):=\textnormal{FFT}(A_i(\mathbf{x});\mathbf{q})$;
  \item Set $A_i(\mathbf{q}):=G_{ij}^0(\mathbf{q})A_j(\mathbf{q})$ for $\mathbf{q}\neq 0$
        and $A_i(\mathbf{q}=0):=\overline{E}_i$;
  \item Set $A_i(\mathbf{x}):=\textnormal{FFT}^{-1}(A_i(\mathbf{x});\mathbf{q})$;
  \item Compute convergence criterion; if convergence is reached, set $E_i=A_i$ and STOP; otherwise GOTO (i).
\end{enumerate}
In this algorithm FFTs are computed \emph{in-place}. Step (iii) consists of a loop over all modes $\mathbf{q}$ with $G_{ij}^0(\mathbf{q})$ computed on-the-fly. In total, memory space is allocated for one vector field $\mathbf{A}$ plus the microstructure. Vector $\mathbf{A}$ successively stores the polarization field in the real space [step (i)] and in the Fourier domain [step (ii)] and the electric field $\mathbf{E}$ in the Fourier domain [step (iii)] and real space [step (iv)]. The convergence criterion in step (v) must be modified, as
checking for current conservation by computing criterion $\eta_1$ with in-place computations is now impractical. Monitoring the differences over two iterations of the first and second moments of the electric and current fields provides practical crieria that are less accurate, but easier to compute.

On the other hand, the use of the discrete Green operator $\widetilde{\mathbb{G}}^0$ allows for a more efficient implementation of the DS. Consider the rewriting of Equation~(\ref{eq:ls1imp}) as
\begin{equation}
\label{eq:dsimp}
\phi^{k+1}=\frac{1}{\sigma^0\Delta}\textnormal{\textbf{div}}\left[(\sigma-\sigma^0)(\mathbf{\overline{E}}-\textnormal{\textbf{grad}}\phi^k)\right]
\end{equation}
where $\phi^k$ is the periodic part of the potential associated to $\mathbf{E}^k$, so that $\mathbf{E}^k-\overline{\mathbf{E}}=-\textnormal{grad}\,\phi^k$, and where $1/\Delta$ is, symbolically, the inverse Laplacian. Equation (\ref{eq:dsimp}) defines $\phi^{k+1}$ as a unique periodic function up to an irrelevant constant. When $k\to\infty$, $\phi^k$ converges to the potential up to a linear correction $\phi^\infty(\mathbf{x})=\Phi(\mathbf{x})-\overline{E}_ix_i$. The electric field and current follow from $\Phi$. In the discrete setting, equivalent to a resistor network, knowledge of a field on adjacent nodes or bonds is sufficient to compute its local divergence or gradient. Thus, the action of the $\textnormal{\textbf{div}}$ and $\textnormal{\textbf{grad}}$ operators in Equation~(\ref{eq:dsimp}) can be computed in the real space. This suggests the following alternative implementation of the discrete direct scheme (\D{DS}):

Initialization: set $A(\mathbf{x})\equiv 0$.
\begin{enumerate}
  \item At each point $\mathbf{x}$, set $A(\mathbf{x}):=\textnormal{\textbf{div}}\,\mathbf{P}(\mathbf{x})$
        where \\
        $P_i(\mathbf{x})=[\sigma(\mathbf{x})-\sigma^0][\overline{E}-\textnormal{\textbf{grad}}\, A(\mathbf{x})]$;
        compute $\eta_1$ as defined in ($\ref{eq:precision}$);
  \item Set $A(\mathbf{q}):=FFT\lbrace A(\mathbf{x});\mathbf{q}\rbrace$;
  \item Set $A(\mathbf{q}):=-\frac{A(\mathbf{q})}{\sigma^0|k(\mathbf{q})|^2}$ for $\mathbf{q}\neq 0$ and $A(\mathbf{q}=0):=0$ otherwise;
  \item Set $A(\mathbf{x}):=FFT^{-1}\lbrace A(\mathbf{q});\mathbf{x}\rbrace$;
  \item If $\eta_1<\epsilon$ set $\mathbf{E}=\overline{\mathbf{E}}-\textnormal{\textbf{grad}}\, A$, $\mathbf{J}=\sigma \mathbf{E}$ and STOP; otherwise GOTO (i).
\end{enumerate}
This algorithm exactly implements the \D{DS} scheme. However, only a scalar field, rather than a vector field, is now allocated in memory. Laplacian inversion is the sole computation performed in the Fourier domain. It takes the form of a division by $|k|^2$ in step (iii).
The field $A$ successively stores the divergence of the polarization field $\textnormal{\textbf{div}}\,\mathbf{P}$ in the real space [step (i)] and Fourier domain [step (ii)] and, later on, the periodic part of the potential $\phi$ in the Fourier domain [step (iii)] and in the real space [step (iv)]. Multithreading parallelization in step (i) necessitates some care as this step is non-local. Nevertheless, this new implementation reduces the number of FFTs per iteration from $4$ (in 2D) or $6$ (in 3D) down to $2$. Furthermore, the amount of storage is also reduced by a factor $d$ ($L^d$ floats instead of $dL^d$), if we neglect the storage required for the microstructure.

\begin{figure}[!ht]
\begin{center}
\vspace{0.5cm}
\includegraphics[width=11cm]{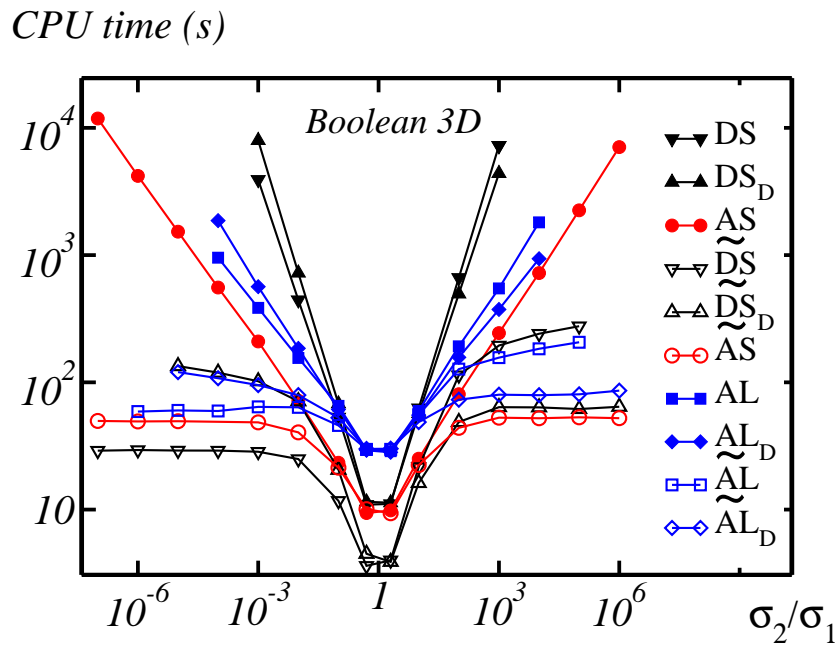}
\caption{\label{fig:cputime} CPU time vs. contrast ratio $\sigma_2/\sigma_1$ for
various FFT algorithms, on the 3D Boolean microstructure.
The convergence criterion is $\eta\leq\epsilon$ with $\epsilon=10^{-8}$.}
\end{center}
\end{figure}

The total CPU time spent using `direct', `augmented-Lagrangian' and `accelerated' schemes is plotted in Figure~\ref{fig:cputime} as a function of contrast, the scheme (\D{DS}) being implemented as outlined earlier. These tests were carried out with convergence criterion $\eta<10^{-8}$, on the previously considered 3D Boolean microstructure discretized on a grid of size $L=256$ (16.8 million points). Computations were performed in double precision, on a $12$-core Intel Xeon machine, each core running at $2.90$ GHz with $5800$ bogomips and $15360$ Kb of L2 cache. Best performance is achieved for the \D{DS} scheme when $\sigma_2/\sigma_1<1$, and with \D{AS} when $\sigma_2/\sigma_1>1$. Using these optimal schemes at infinite contrast, convergence is completed in $29$ s for insulating inclusions, and in $53$ s for infinitely-conducting inclusions. This strategy has been implemented in the multithreaded Fortran code \url{morph-hom} developped at Mines ParisTech~\cite{morphhom}.
% ----------------------------------------
% ----------------------------------------

\section{Conclusion}\label{sec:conclusion}
Use of a modified Green operator in FFT-based schemes has been advocated, in the context of the electrical response of heterogeneous conducting media. The modification consists in making the operator consistent with the underlying voxel grid, which requires only a very simple adaptation of previously existing algorithms but leads to two major improvements.

First, employing the modified operator leads to much more accurate local fields, particularly in the highly conducting or insulating inclusions and in the vicinity of interfaces. Second, the convergence rate is found to be much faster compared with previous methods, in particular for highly-contrasted media. Quite remarkably the `direct' scheme ---usually considered to be the worst-converging one--- improves tremendously, as far as CPU time is concerned, by formulating the problem in terms of iterations on the electrostatic potential rather than on the electric field. However, using the modified Green operator requires carefully adjusting the reference conductivity $\sigma^0$, since the latter has a strong influence on convergence properties. Approximate expressions for $\sigma^0$ have been derived numerically, and studied, for the Boolean models of microstructure considered in this work.

It has already been noticed in the past that using `discrete' versions of Green operators leads to promising methods~\cite{Brisard10,wiegmann2006,willot08c}. Demonstrating that dramatic speed-up improvements follow, the present work strongly supports this view. Based on previous experience~\cite{willot08c}, it is expected that our conclusions carry over to continuum mechanics.

% ----------------------------------------
\ack
The authors are grateful to H.\ Moulinec for kindly providing some field maps for comparison purposes, which has been a very helpful assistance in this study. The research leading to the results presented has received funding from the European Union's Seventh Framework Programme (FP7 / 2007-2013) for the Fuel Cells and Hydrogen Joint Technology Initiative under grant agreement 303429.
% ----------------------------------------

\end{document}